# Video Pandemics: Worldwide Viral Spreading of Psy's Gangnam Style Video

Zsófia Kallus[1,2], Dániel Kondor[1,3], József Stéger[1], István Csabai[1], Eszter Bokányi[1], and Gábor Vattay[1](✉)

[1] Department of Physics of Complex Systems, Eötvös University
Pázmány P. s. 1/A, H-1117, Budapest, Hungary
vattay@elte.hu
[2] Ericsson Research, Budapest, Hungary
[3] Senseable City Lab, Massachusetts Institute of Technology, Cambridge, USA

**Abstract.** Viral videos can reach global penetration traveling through international channels of communication similarly to real diseases starting from a well-localized source. In past centuries, disease fronts propagated in a concentric spatial fashion from the the source of the outbreak via the short range human contact network. The emergence of long-distance air-travel changed these ancient patterns. However, recently, Brockmann and Helbing have shown that concentric propagation waves can be reinstated if propagation time and distance is measured in the flight-time and travel volume weighted underlying air-travel network. Here, we adopt this method for the analysis of viral meme propagation in Twitter messages, and define a similar weighted network distance in the communication network connecting countries and states of the World. We recover a wave-like behavior on average and assess the randomizing effect of non-locality of spreading. We show that similar result can be recovered from Google Trends data as well.

## 1 Introduction

According to Wikipedia, the music video of 'Gangnam Style' by recording artist Psy reached the unprecedented milestone of one billion YouTube views on December 21, 2012. It was directed by Cho Soo-Hyun and the video premiered on July 15, 2012. What makes this viral video unique from the point of view of social network research is that it was spreading mostly via human-to-human social network links before its first public appearance in the United States on August 20, 2012 and Katy Perry shared the music video with her 25 million followers on Twitter on August 21.

Other viral videos spread typically via news media and reach worldwide audiences quickly, within 1-3 days. Our assumption is that only those online viral phenomena can show similarities to global pandemics that were originally constrained to a well localized, limited region and then, after an outbreak period, reached a worldwide level of penetration. In 2012, the record breaking 'Gangnam Style' [1] marked the appearance of a new type of online meme, reaching unprecedented level of fame despite its originally small local audience. From the sub-culture of k-pop fans it reached an increasingly wide range of users of online media - including academics - from around the World (Refs. [2–5]). We approximately reconstructed its spreading process by filtering geo-tagged messages containing the words 'Gangnam' and 'style'. In Fig.1 we show the location of geo-tagged posts containing the expression 'Gangnam Style' from our collection of the public Twitter feed, as of September 2012. For this purpose we used our historical Twitter dataset and collection of follower relations connecting 5.8M active Twitter users who enabled access to their location information while posting messages to their public accounts (see Ref. [6] for details). When tracing videos on the Twitter social platform we have access only to the public posts, and only look at geo-tagged messages. Location information allows us to record the approximate arrival time of a certain news to a specific geo-political region. In the real space this process looks indeed random, but the 'local to global' transition is also apparent as the messages cover a progressively larger territory. We collected the approximate first arrival time of the video in different geo-political regions of the World. In order to study the viral spreading of the appearances of the video in the Twitter data stream first we coarse grained the World map into large homogeneous geo-political regions. We used regions of countries and states of the World as the cells (i.e. the nodes) and aggregated the individual links connecting them. By performing an aggregation into geo-political regions we construct a weighted graph connecting 261 super nodes. Thus the ratio of edge weights can be interpreted as an approximation of the relative strength of communication between pairs of the connected regions. This high-level graph is thus naturally embedded into the geographic space giving a natural length to its edges. We then connect individuals in the spatial social network and then recreate a high-level aggregated weighted graph between regions by querying a large database containing the collection of historic, freely available Twitter messages [6–8]. In Fig. 2 we show the resulting weights in graphical form for Twitter users in California.

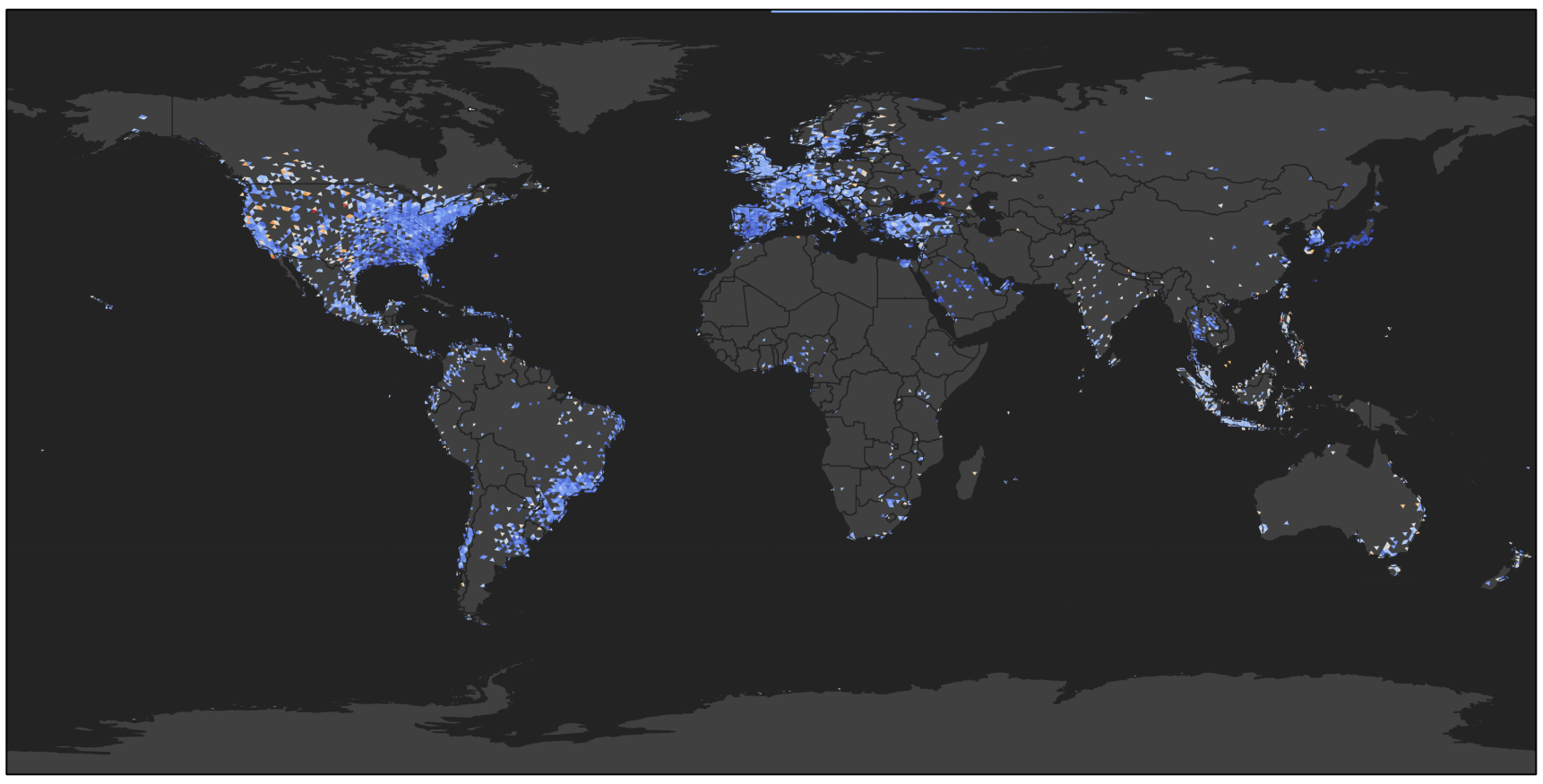

**Fig. 1. Geo-locations of Twitter messages containing 'Gangnam Style'.**

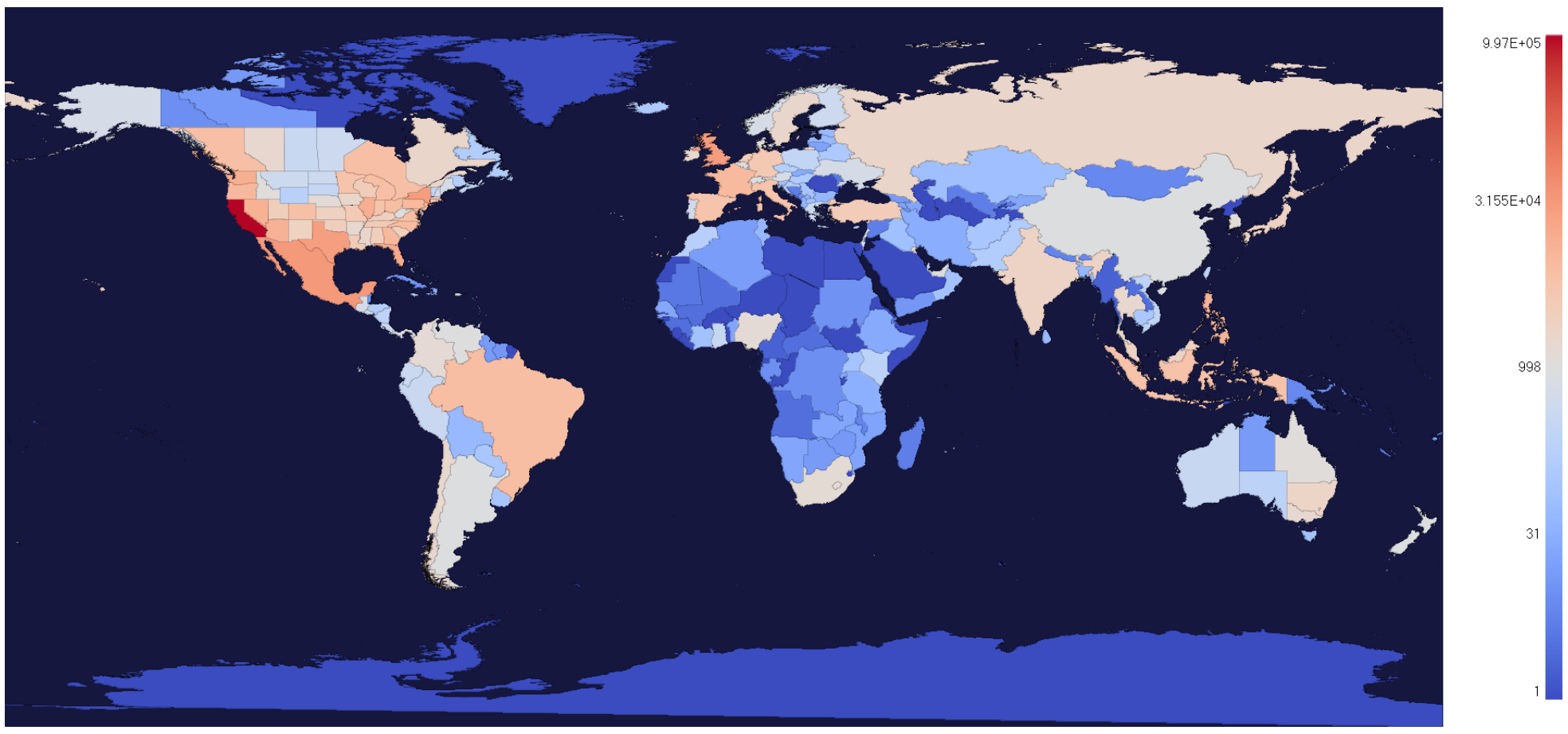

**Fig. 2. Social connection weights between large geo-political regions of the World.** The map shows our 261 geo-political regions and the number of friendships (mutual Twitter followers) between users in California and the rest of the World. Colour codes the number of friendships with users in California in our database. Deep red means that Californians have $\sim 10^5$ friendships within California and deep blue indicates that $\sim 10^0$ friendship connects them for example to certain regions of Africa.

## 2 Weighting the speed of spreading in the network

Brockmann and Helbing in Ref. [9] worked out a method by which they were able to predict the effective spreading time of a real disease between two nodes of air-traffic based on the network and the number of passengers travelling between them in unit time and introduced an effective measure of distance of the two nodes. Here, we repeat their derivation, except, we use the number of mutual social contacts (mutual Twitter followers called friends) between geo-located Twitter users in the geo-political regions. Each user can be in one of two possible states: *susceptible* for the viral video (never seen the video before) or *infected* (already seen the video) and affects the state of others when contact occurs between them via sharing the video in their social network feeds. The model that we adopt is based on the meta-population model [10,11]. After dividing the world map into geo-political regions the dynamics within the $n^{th}$ spatial unit is modelled by the *SI equations*( [12–15]) with disease specific parameters:

$$\begin{aligned} \partial_t S_n &= -\alpha I_n S_n / N_n, \\ \partial_t I_n &= -\beta I_n + \alpha I_n S_n / N_n \ . \end{aligned} \tag{1}$$

This local dynamics is complemented by the weights that connect the separated cells as described by

$$\sum_{m \neq n} w_{mn} U_m - w_{mn} U_n. \tag{2}$$

Here $U_n$ represents the $n$th *SI* state variable, and $w_{mn} = F_{nm}/N_m$ is the per capita traffic flux from site $m$ to site $n$, $F_{nm}$ being the weighted adjacency matrix representing the network. This means that the human contact network can be effectively divided into interconnected layers and the spatial reach of a spreading process is determined mainly by the weighted network of regions. We only need this high-level information and the details of the large and complex temporal contact network can be neglected. Once we have a weighted graph and the arrival times of the video at each of the nodes the embedding of the graph into an abstract space can be performed. Its goal is to uncover the wave pattern of the dynamics. This means finding a source region from which the dependence of the arrival times at a region is linear on the effective distance of the region from the origin i.e., there exists an effective velocity. If the spreading is governed by the equations (1-2) the effective distance (spreading time) between nodes $m$ and $n$ can be defined as follows:

$$d_{mn} = 1 - log(P_{mn}) \leq 1, \tag{3}$$

where $P_{mn} = F_{mn}/\sum_m F_{mn}$ is the flux fraction from $m$ to $n$ i.e., the probability to choose destination $m$ if one is in the region $n$. This way a minimal distance means maximal probability, and additivity of the distances is ensured by the logarithmic function.

## 3 From geographic to network-based embedding

The embedding comprises of the following steps. After calculating the connection weighted adjacency matrix, each link weight is transformed into the *effective quasi distance* defined by Eq. 3. Starting from each node a *shortest path tree* can be constructed from all nodes reachable from the selected one. These shortest paths correspond by definition to the most probable active routes that the equations of epidemics and the weights would predict. In such a tree the distance of each node from the origin is the length of the shortest path connecting them. The original graph and one of its shortest path trees is shown in Fig. 3. If the node selected as origin corresponds to the most effective source, the video will arrive first at the closest nodes and then propagate towards the periphery. If the arrival times show a linear dependence on the effective distance, it is likely that we found the source of the spreading. Comparing different trees by measuring the goodness of a linear fit allows us to find the most probable and most effective source node on the graph. For the ‘Gangnam Style’ video it is the region of the *Philippines*. While the obvious center should be Seoul/South Korea, where the video has been created, it seems that the most intensive source of social network spreading was in the Philippines. This is probably due to the fact that we are not able to measure the very short time it took the video to spread from South Korea to the Philippines and more importantly, the Philippines is much more connected socially to the rest of the world than South Korea. This is partially due to the English language use and a well spread diaspora of the Philippines.

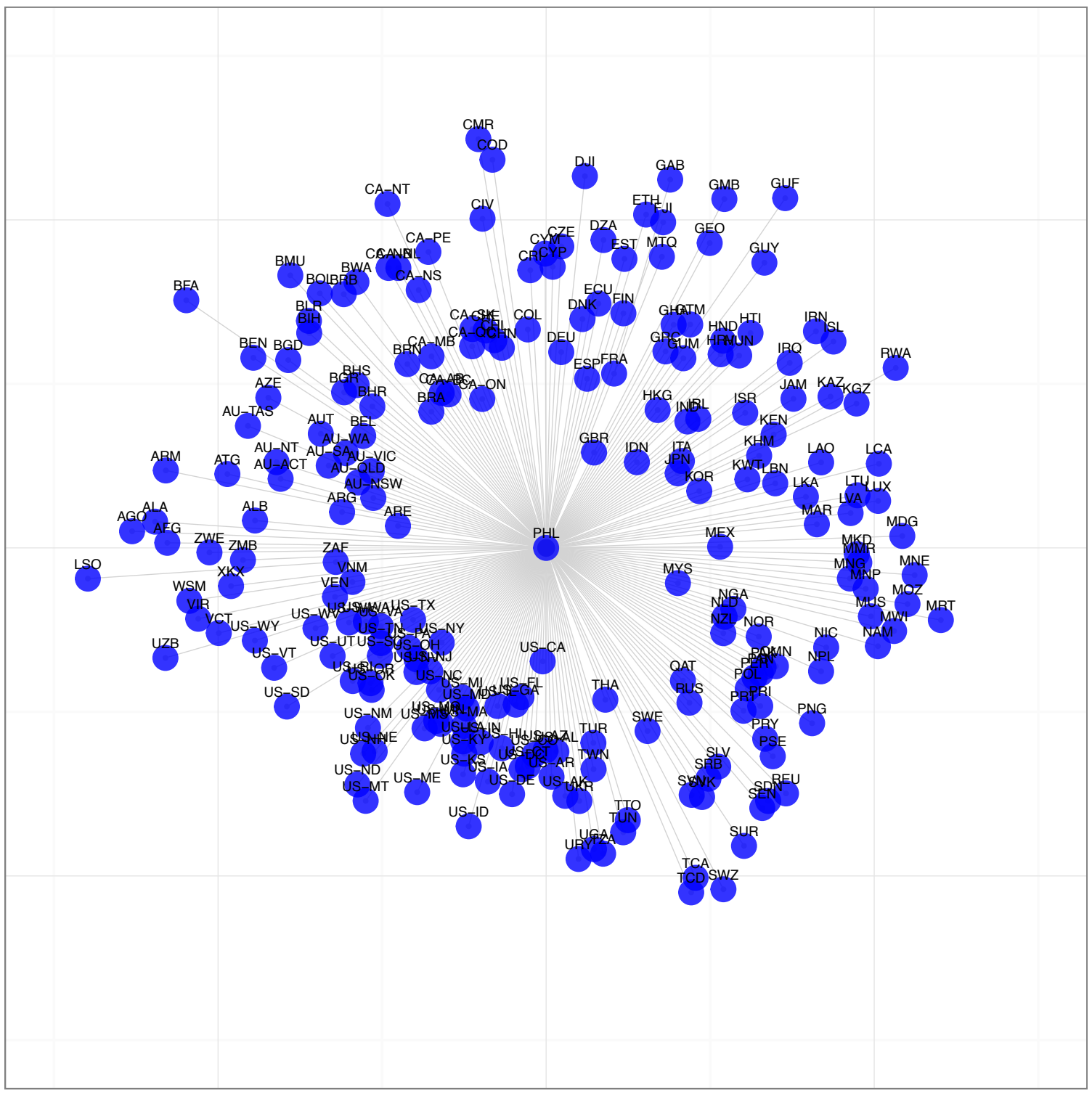

**Fig. 3. Embedded shortest path tree.** The countries of the World are partitioned to 261 large administrative regions. An aggregated version of the weighted and directed graph has been used from the individual-level follower relations. Effective distances are represented as the radial distance from the node at the origin in arbitrary units, and the angular coordinate of the nodes is arbitrary as well.

The linear fit, on the other hand, is not perfect. Uncertainty is introduced into our analysis by multiple factors. These are the heterogeneous nature of the use of the social platforms around the world and the variability of activities over time; the partial measurements obtained from the publicly available geo-tagged sample of tweets; and the external effect of other media propagating the same news. These circumstances all add to the uncertainty of the first arrival times in our dataset. In order to recover the average wave form we had to use smoothing in space and in time as well. The spatial averaging achieved by regional aggregation is also justified by the assumption that users within the same region are likely to be more connected to each other than to the rest of the world [8]. We used a moving window over the arrival times and used the linear fitting to the average distances and the average times of the windows. The noise can be effectively reduced by choosing a window of two weeks to a month. As shown on Fig. 4 by the colour scale, the smaller the number of Twitter geo-users in a region, the less reliable the analysis becomes. Moreover these small regions form the peripheral ring as seen from the most probable source region's point of view.

## 4 Comparison of Twitter and Google Trends distances

Once the site of origin is selected, the embedding is straight-forward. The effective distances between nodes and the source node is equivalent to the shortest path distances on the embedded graph. Fig. 5 shows how the underlying order can be uncovered by this transformation. The seemingly randomized left panel – representing the arrival of the video at various *geographic* distances – becomes structured, and a linear trend emerges as a function of the *effective* distance. Linearity breaks down only at large distances, where remote peripheral regions are left waiting for the video to arrive at last. We also measured arrival times by looking at the Google search engine records through Google Trends analytics service. In Figure6 we show the results. Its sparsity is coming from a lower temporal resolution of Google Trends compared to our Twitter based dataset. It is, however, very remarkable that the network embedding we found for the Twitter data works also for the Google Trends dataset and creates the proper reordering of the nodes. This result further underlines the effectiveness of our Twitter measurements.

In order to check the consistency of our results with other sources we analyzed an additional set for the regional arrival times. For this purpose we used publicly available service of the Google Trends platform, and gathered the cumulative number of historic web searches performed with the keywords *Gangnam* and *style*. We were able to create a dataset with a resolution of 2 weeks. As shown on the second panel of Fig. 6 the linear dependence is once again recovered, and the linear propagation of the wave can be effectively seen. This proves that Twitter predicts correctly the flow of news and information on the global network of communication. Searches performed first in each region seem to be synchronized with the arrival times of the first tweets coming from Twitter.

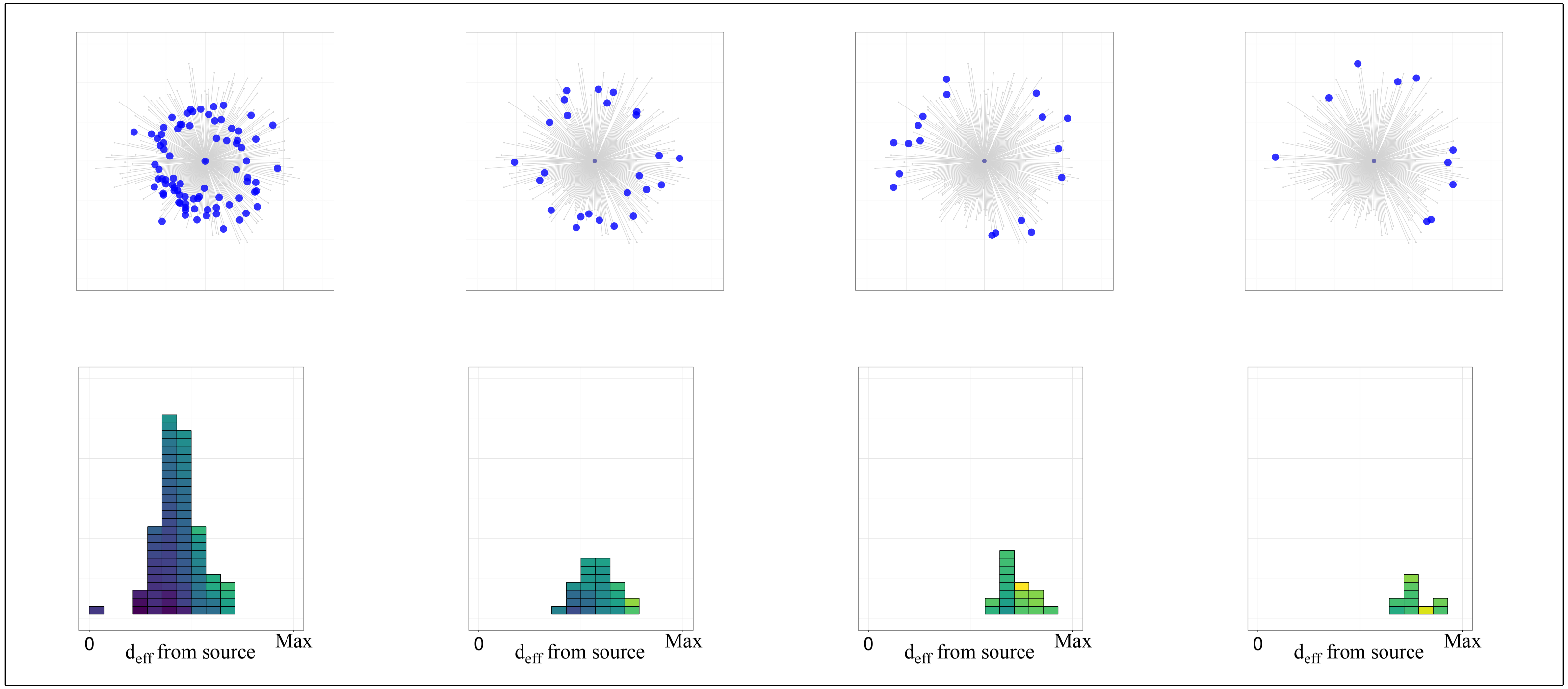

**Fig. 4. Progressive stages of the pandemic.** The spreading of the wave is shown in four progressive stages of the propagation. Each stage is defined by separate time slice of equal length. The nodes where the news has just arrived in that slice are first shown on the shortest path tree. Second, a corresponding histogram is created based on effective distances. Each rectangle represents one of the regional nodes and a common logarithmic color scale represents the number of users of the nodes (color scale of Fig. 5 is used).

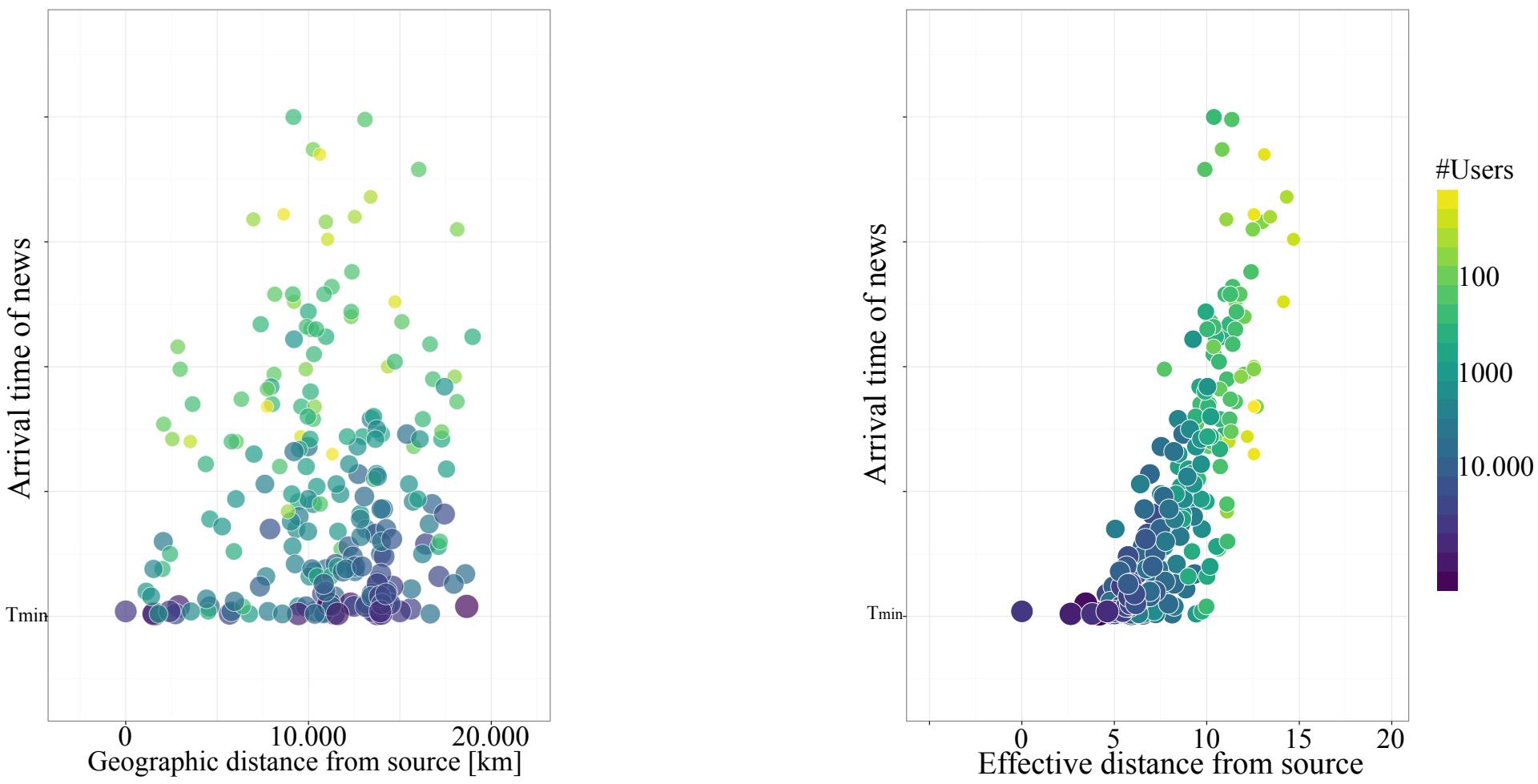

**Fig. 5. Geographic distance vs effective distance on Twitter.** Here each dot represents a state, colored according to the number of users, using the same logarithmic color scale as before. The horizontal axis represents the distance from the source node of the Philippines, while the vertical axis is the arrival time of the video at that regional node. The clear order is uncovered in the second panel, where geographic distance is replaced by the effective distance.

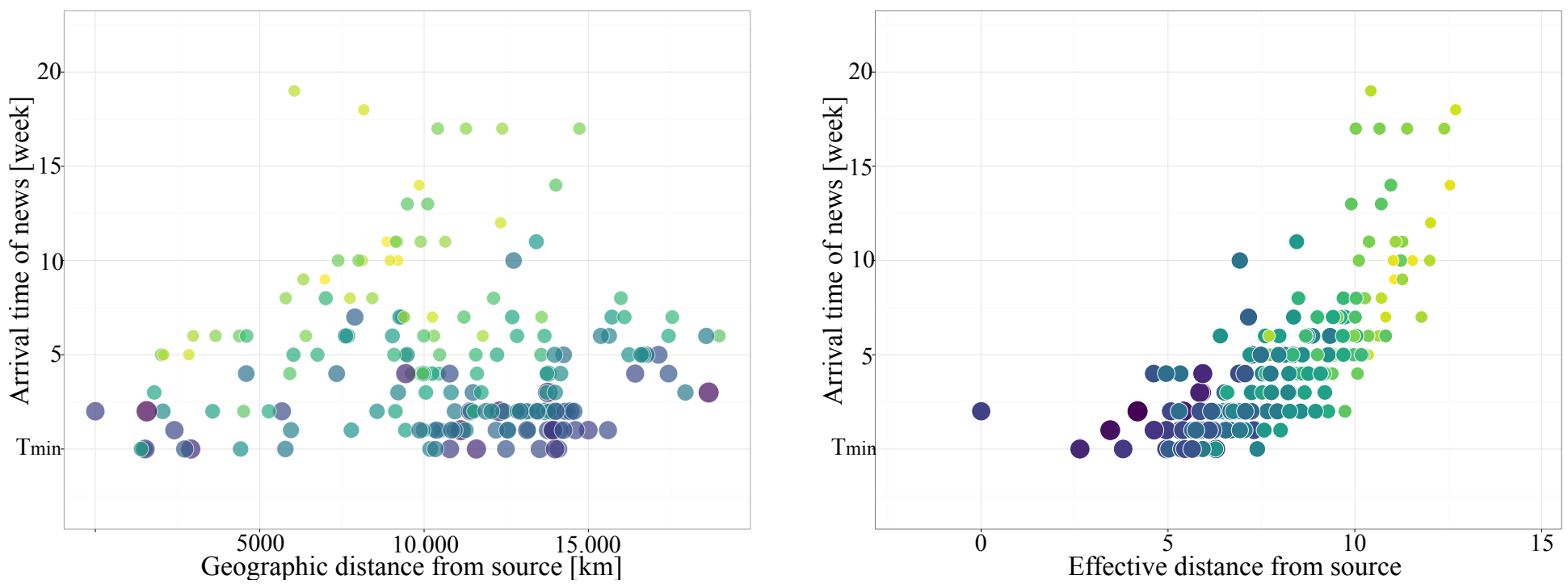

**Fig. 6. Geographic distance vs effective distance in Google Trends.** Here each dot represents a state, colored according to the number of Users (same color scale as Fig. 5), using the same logarithmic color scale as before. The horizontal axis represents the distance from the source node of the Philippines, while the vertical axis is the arrival time of the video at that regional node. The clear order is uncovered in the second panel, where geographic distance is replaced by the effective distance.

## 5 Conclusions

In this paper we investigated the online spreading of the viral video 'Gangnam Style' in the social network Twitter. Using geo-localized tweets we determined the first appearance of the video in the 261 major geo-political areas of the World. We adapted the tools developed in Ref. [9] for the spreading of infectious diseases over the air-travel network for the present problem. We showed that the number of friendships (mutual following) of Twitter users between geo-political regions is analogous to the passenger traffic volume of air-traffic from the point of view of online information spreading. Using our historic public tweet database we could calculate the relative weights of information traffic between the regions. We then reproduced the shortest path effective distance of geo-political regions from the epicenter of the social outbreak of the 'Gangnam Style' video pandemic, which seems to be in the Philippines, not far from South Korea, where it has been produced. We managed to verify the spreading pattern independently, based on the search results in Google Trends in the same time period. The synchrony between first appearances in Twitter and Google suggest that a universal pattern of social information flow (information highways) exist between geo-political regions, which is inherently non-technological, determined by the strength of social ties between different countries, cultures and languages. Further research is necessary to understand the main features of this global network.

**Acknowledgement.** The authors would like to thank János Szüle, László Dobos, Tamás Hanyecz, and Tamás Sebők for the maintenance of the Twitter database at Eötvös University. Authors thank the Hungarian National Research, Development and Innovation Office under Grant No. 125280 and the grant of Ericsson Ltd.